\documentclass[journal]{IEEEtran}
 \usepackage{amsmath,amssymb}
 \usepackage{subfigure}
 \usepackage{graphicx,graphics,color,psfrag}
 \usepackage{cite,balance}
 \usepackage{caption}
 \allowdisplaybreaks
 \usepackage{algorithm}
 \usepackage{accents}
 \usepackage{amsthm}
 \usepackage{bm}
 \usepackage{algorithmic}
 \usepackage[english]{babel}
 \usepackage{multirow}
 \usepackage{enumerate}
 \usepackage{cases}
 \usepackage{stfloats}
 \usepackage{dsfont}
 \usepackage{color,soul}
 \usepackage{amsfonts}
 \usepackage{cite,graphicx,amsmath,amssymb}
 \usepackage{subfigure}
 \usepackage{fancyhdr}
 \usepackage{hhline}
 \usepackage{graphicx,graphics}
 \usepackage{array,color}
 \usepackage{amsmath}
 \usepackage{booktabs}

\def\phi{\varphi}

\def\({\left(}
\def\){\right)}

\def\b0{{\mathbf{0}}}

\begin{document}
\setlength{\topskip}{-3pt}

\title{\huge Over-the-Air Integrated Sensing, Communication, and Computation in IoT Networks}
\author{Xiaoyang Li, Yi Gong, Kaibin Huang, and Zhisheng Niu
\thanks{Xiaoyang Li and Yi Gong are with the the Department of Electrical and Electronic Engineering (EEE), Southern University of Science and Technology (SUSTech), Shenzhen, China. Kaibin Huang is with the Department of EEE, The University of Hong Kong, Hong Kong. Zhisheng Niu is with the Department of EEE, Tsinghua University, Beijing, China. Corresponding author: Yi Gong (e-mail: gongy@sustech.edu.cn).} 
}
\maketitle

\begin{abstract}
To facilitate the development of Internet of Things (IoT) services, tremendous IoT devices are deployed in the wireless network to collect and pass data to the server for further processing. Aiming at improving the data sensing and delivering efficiency, the \emph{integrated sensing and communication} (ISAC) technique has been proposed to design dual-functional signals for both radar sensing and data communication. To accelerate the data processing, the function computation via signal transmission is enabled by \emph{over-the-air computation} (AirComp), which is based on the analog-wave addition property in a multi-access channel. As a natural combination, the emerging technology namely \emph{over-the-air integrated sensing, communication, and computation (Air-ISCC) adopts both the promising performances of ISAC and AirComp to improve the spectrum efficiency and reduce latency by enabling simultaneous sensing, communication, and computation.} In this article, we provide a promptly overview of Air-ISCC by introducing the fundamentals, discussing the advanced techniques, and identifying the applications.
\end{abstract}

\section{Introduction}
Prompted by the next-generation wireless networks (6G and beyond), a series of advanced \emph{Internet of Things} (IoT) services have been conducted, such as \emph{artificial intelligence} (AI), autonomous driving, smart cities, virtual reality, and digital twins \cite{saad2019vision}. Towards this vision, tremendous IoT devices are expected to collect and deliver data from the environment to the server for further fusion and processing \cite{cui2021integrating}. In conventional schemes, the sensing, communication, and computation processes are separately designed, which results in not only high cost but also low efficiency \cite{liu2020joint}. By designing dual-functional signals for both data communication and radar sensing, \emph{integrated sensing and communication} (ISAC) has been proposed to improve the data sensing and delivering efficiency \cite{liu2022integrated}.

%Nevertheless, the computation part remains isolated as it mainly lies beyond the physical layers.To improve the efficiency, the operations of sensing, communication, and computation are expected to be integrated together.  and gradients aggregation in federated learning

Meanwhile, as many IoT applications only require the statistical information (e.g., distributed consensus in fleet driving) rather than the data symbols themselves, accurate function computation is more expected than raw data delivery. To this end, \emph{over-the-air computation} (AirComp) enables the function computation via signal transmission, which is based on the analog-wave addition property in a \emph{multi-access channel} (MAC) \cite{zhu2021over}. The main idea of AirComp is to exploit interference for computing functions, and thus making the wireless channel to be a computer \cite{nazer2007computation}. As a natural combination of ISAC and AirComp, the emerging technology namely \emph{over-the-air integrated sensing, communication, and computation} (Air-ISCC) enables simultaneous sensing, communication, and computation via proper radar signal design.

%A natural tradeoff exists between the performance of the two functionalities, which is characterized as the \emph{mean squared errors} (MSE) of target estimation for radar sensing and the MSE of function computation for AirComp. To tackle the challenges that Air-ISCC might face in the practical implementation, a series of researches focus on exploring diverse directions including waveform design, spatial multiplexing, and target estimation.

\newpage

In Air-ISCC, the radar signals transmitted by multiple IoT devices both detect the targets and carry data symbols, while the server receives the statistical information of data symbols via AirComp \cite{li2022ISAA}. The emerging Air-ISCC technologies provide a promising vision for supporting more advanced IoT functionalities, which will facilitate the evolution of wireless networks from “connected things” to “connected intelligence” \cite{letaief2019roadmap}. Compared with the conventional approaches, Air-ISCC is expected to collect and process data from more devices within less latency via proper signal design. In application scenarios requiring massive machine type communication (mMTC), devices connected via conventional multiplexing methods compete for the radio resource fiercely, while Air-ISCC avoids such competition by sharing the same spectrum for concurrent sensing and multi-user communication. In application scenarios requiring ultra-reliable low-latency communications (URLLC), Air-ISCC can significantly reduce the latency by performing sensing, communication, and computation simultaneously.

In this article, we provide an overview of Air-ISCC by introducing the fundamentals in Section II, discussing the advanced techniques in Section III, and identifying the possible applications in Section IV. It should be noted that despite the existing literatures about ISCC \cite{qi2020integrated,shi2021integration}, this is the first overview paper of the Air-ISCC to the best of the authors' knowledge.

\begin{center}
\noindent \fbox{\parbox{.9\linewidth} {\center \textbf{Guideline of Air-ISCC:} \\
--------------------------------------------------------------------\\
Enable simultaneous sensing, communication, and computation by proper radar signal design to improve spectrum efficiency and reduce latency.}}
\end{center}

%In the Air-ISCC framework as shown in Fig.~\ref{FigSys}, an IoT network is designed to enable the simultaneous sensing, communication, and computation functionalities. In particular, multiple IoT sensors transmit dual-functional signals for targets detection and data delivery to servers via AirComp. 

\begin{figure*}[ht]
\centering
\includegraphics[scale=0.45]{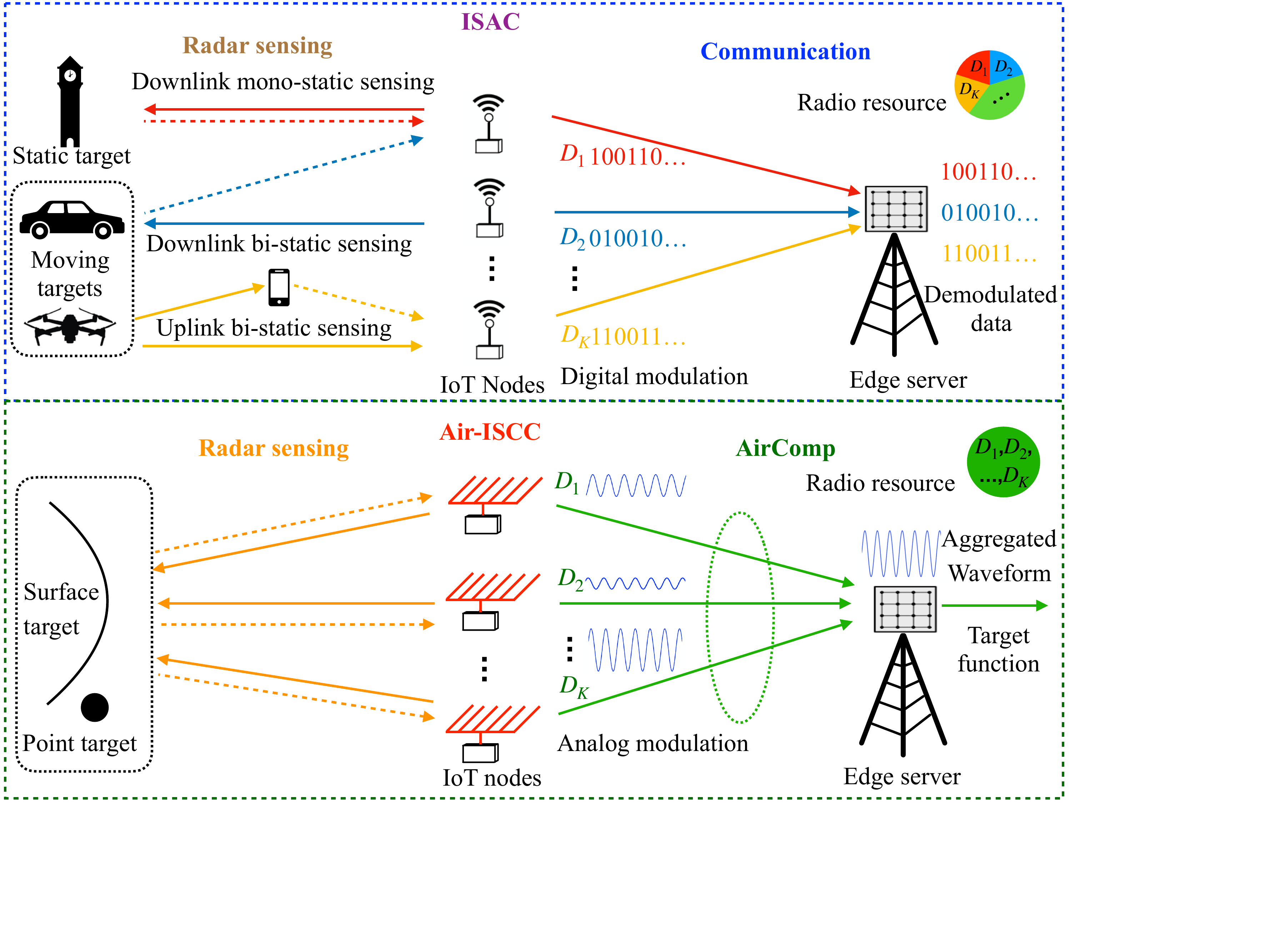}
\caption{Paradigms comparison: ISAC versus Air-ISCC.}
\label{FigSys}
\end{figure*}

\section{Air-ISCC Fundamentals}
\subsection{AirComp}
The fundamental idea of AirComp lies in exploiting the analog-wave addition property in MAC. Based on such mechanism, the 
signals simultaneously transmitted by IoT devices are superposed over-the-air and aggregated at the server with the form of weighted summation, where the weights represent the channel conditions \cite{zhu2021over}. To support AirComp, channel pre-compensation and linear-analog modulation at each transmitter are necessary. In channel pre-compensation, the heterogeneous channels are equalized to guarantee the accuracy of AirComp. In linear-analog modulation, the data values are modulated into the magnitudes of the carrier signals. After such process, the received signal is a vector composed of the transmitted data scaled by a factor for balancing the noise and misalignment. By adjusting the factor, the average of the distributed data can be recovered from the aggregated signal and thus realizes the AirComp. AirComp can also be digitalized by adopting a multi-level linear analog modulator with complex amplitude quantization.

Besides the simple average, AirComp can compute more complex functions via proper data pre/post-processing, which refers to a post-processed summation of multiple pre-processed data-values. These functions are known as the nomographic functions, including weighted sum, arithmetic mean, polynomial, geometric mean, and Euclidean norm. It has been further proved that any function can be approximated expressed by a series of nomographic functions, and thus can be computed via AirComp in general. To support multi-functions computation simultaneously, the concepts of \emph{multi-modal sensing} together with the antenna arrays are utilized to enable the \emph{multiple input multiple output} (MIMO) AirComp \cite{li2019wirelessly}. The promising performance of AirComp in fast data aggregation makes it popular in a series of advanced IoT applications, such as the distributed consensus \cite{zhu2021over} and the edge learning \cite{zhu2020toward}.

%\begin{table}[ht]
%\centering
%\caption{Summarization of nomographic functions.}
%\begin{tabular}{|p{4cm}|p{4cm}|}
%\hline
%\bf{Name} & \bf{Expression} \\
%\hline
%General Form &  $y = f\l(\sum_{k=1}^{K}g_{k}(x_k)\r)$ \\ 
%\hline
%Weighted Sum &  $y = \sum_{k=1}^K \omega_k x_k$ \\ 
%\hline
%Arithmetic Mean &  $y = \frac{1}{K}\sum_{k=1}^K x_k$ \\ 
%\hline
%Polynomial &  $y = \sum_{k=1}^K \omega_k x_k^{\beta_k}$ \\ 
%\hline
%Geometric Mean &  $y = \l(\prod_{k=1}^K x_k \r)^{1/K}$ \\ 
%\hline
%Euclidean Norm &  $y = \sqrt{\sum_{k=1}^K x_k^2}$ \\ 
%\hline
%\end{tabular}
%\label{summary:table1}
%\end{table}

\subsection{ISAC}
In practical spectrum, the S-band (2-4 GHz) and C-band (4-8 GHz) for radar sensing can also be used for communication \cite{cui2021integrating}. A series of researches focus on improving the performance of radar sensing and data communication, including the waveform design, MIMO beamforming, interference cancellation, and so on. As the details of ISAC development are thoroughly reviewed in \cite{liu2020joint} and \cite{liu2022integrated}, a brief summary is given here to avoid duplication. From the perspective of information theory, the rate distortion theory was applied to unify the performance of radar sensing and data transmission. Consequently, a dual-functional waveform design was implemented in single antenna system to support data delivery and target estimation simultaneously. Along this vein, the waveform for supporting both radar sensing and data communication was further extended into the MIMO systems, with the information bits embedded in the sidelobe of the radar signal beampattern. To support communication system with multi-users, a brunch of transmit beamforming designs were investigated. Aiming at reducing signal distortion, the dual-functional beamforming design further took the constant modulus waveforms into consideration. The benefit brought by spectrum sharing makes ISAC a popular technology for a series of IoT systems, such as smart home, \emph{unmanned aerial vehicle} (UAV) systems, edge learning systems, vehicular networks, and \emph{reconfigurable intelligent surface} (RIS) systems. 

Among the rich literature on ISAC, the incorporation of data computation is seldom visited as it mainly lies beyond the physical layers. As an emerging technique, AirComp enables fast function computation via transmissions in the same layer of ISAC. Therefore, these two techniques are naturally combined to facilitate the operations of sensing, communication, and computation.

\subsection{Air-ISCC}
In Air-ISCC system as depicted by Fig.~\ref{FigSys}, multiple IoT devices are expected to detect targets and deliver information to the server, while only the statistical functions (e.g., nomographic functions) rather than the raw data are concerned by the server \cite{li2022ISAA}. From the perspectives of signal design and performance analysis, Air-ISCC is different from ISAC in the following parts. First, instead of digital modulation in ISAC, the signals in Air-ISCC are generated based on linear analog modulators. Second, the communication objective in ISAC is to recover the individual data from each IoT device, which might suffer from the interference caused by other communication links. In contrast, as Air-ISCC aims at receiving the statistical functions with lower \emph{mean squared error} (MSE), the signals from all IoT devices are superposed together in the same frequency band without interference. The above differences make Air-ISCC outperform ISAC in the aspects of spectrum efficiency and data processing latency. However, Air-ISCC suffers from the security and synchronization problems due to the concurrent signals transmission, which might be alleviated by the designs against attack and eavesdropping as well as the asynchronous design for AirComp.

%All the IoT devices can transmit signals simultaneously to detect the target and deliver the information to the server via AirComp. A reference clock broadcast by the server is applied for synchronizing the operations of different IoT devices. Each IoT device individually estimates its \emph{channel state information} (CSI) between the server from the broadcasted pilot signals and then passes it to the server. The \emph{key performance indicator} (KPI) for radar sensing is the \emph{mean squared error} (MSE) of target estimation, while the KPI for AirComp is the MSE of function computation. The interference in radar sensing can be effectively reduced via statistical methods such as the law of large numbers, while the interference in data transmission can be harnessed to help the function computation via AirComp. Therefore, Air-ISCC is expected to be a promising approach for alleviating the shortage of spectrum resource and boosting the development of wireless networks to 6G and beyond.

The researches about Air-ISCC can be generally divided into multiple categories from the different perspectives. Starting from the objectives, one direction is to improve the radar sensing performance while guaranteeing the AirComp accuracy, which is known as the \emph{sensing centric Air-ISCC}, while the contrary direction namely \emph{AirComp centric Air-ISCC} aims at increasing the AirComp accuracy without violating the threshold of radar sensing MSE. From the perspective of \emph{target types}, the target to be estimated can be modeled as a point, a surface, a static object, or a moving object, which will result in different problem formulations and solutions. Moreover, according to the usage of antenna arrays, the relevant waveform and beamforming schemes can be categorized as the \emph{shared or separated design}, where the former utilizes all the antennas to transmit a joint waveform for both radar sensing and AirComp, while the latter splits antenna array for radar sensing and AirComp respectively.

\begin{figure*}[ht]
\centering
\includegraphics[scale=0.45]{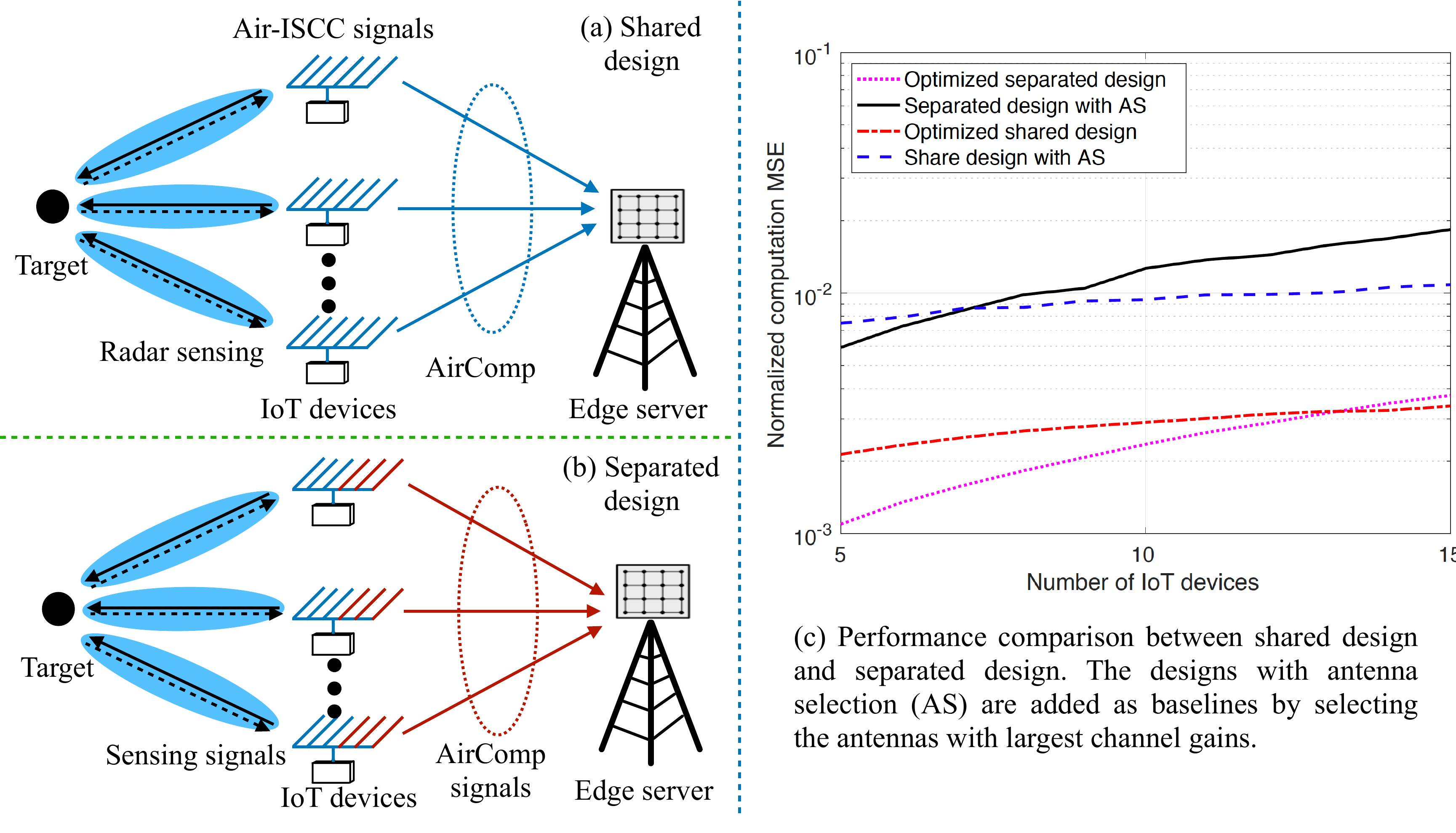}
\caption{Spatial multiplexing: Shared design versus separated design.}
\label{FigSpatial}
\end{figure*}

\section{Advanced Air-ISCC Techniques}
Recent researches on Air-ISCC focus on advanced techniques for reducing radar sensing errors, improving AirComp accuracies, or supporting target detection in practical scenarios. To this end, multiple promising directions in this field warrant further investigation, including spatial multiplexing, target estimation, waveform design, and security design.

\subsection{Spatial Multiplexing for Air-ISCC}
To enable the multi-targets detection and multi-functions computation over-the-air, spatial multiplexing is required to support MIMO Air-ISCC. As illustrated in Fig.~\ref{FigSpatial}, multiple IoT devices equipped with multi-antennas transmit signals for radar sensing and AirComp. The tradeoff between the performances of radar sensing and function computation is reflected in the beamforming design, which necessitates the joint design of transmission beamformer at each IoT device and the data aggregation beamformer at the server. The former aims at balancing the performances of sensing and AirComp, while the latter aims at reducing the AirComp error via noise suppression. The beamforming designs together with the performance analysis are conducted in \cite{li2022ISAA}.

%\begin{center}
%\noindent \fbox{\parbox{.9\linewidth} {\center \textbf{Guideline for Air-ISCC Spatial Multiplexing:} \\
%--------------------------------------------------------------------\\
%Improve the sensing and AirComp performance via beamforming design.}}
%\end{center}

\textbf{Research opportunities:}

\begin{itemize}
\item {\bf Shared antenna beamforming design for Air-ISCC}: As shown in Fig.~\ref{FigSpatial} (a), since the signal transmitted by each IoT device serves both as a radar probing pulse and a data carrier in the shared design, only one beamformer needs to be designed at each IoT device, namely the transmission beamformer. By applying the \emph{maximum likelihood estimation} (MLE), the information of the target can be estimated from the reflected signal. To support AirComp, a data aggregation beamformer at the server is deployed for equalizing the channels of the signals. The joint transmission and data aggregation beamforming design can be formulated as a semidefinite programming problem for minimizing the AirComp error while guaranteeing the radar sensing requirement and power budget for each IoT device. The solving approach based on \emph{semidefinite relaxation} (SDR) is applied to obtain a tractable solution, while the optimal solution warrants further investigation.

\item {\bf Separated antenna beamforming design for Air-ISCC}: As shown in Fig.~\ref{FigSpatial} (b), the antennas at each IoT device are split to transmit one signal for data transmission and another for radar sensing. Therefore, there are two beamformers to be designed at each IoT device, which are known as the radar sensing beamformer and the data transmission beamformer, respectively. The existence of radar sensing signal will result in extra interference for AirComp at the server. The coupling between data transmission beamformer, radar sensing beamformer, and data aggregation beamformer makes the optimization problem more challenging to be coped with.
\end{itemize}

Fig.~\ref{FigSpatial} (c) illustrates the initial performance comparison between the shared and separated designs. The \emph{antenna selection} (AS) scheme is added as a baseline by selecting particular numbers of antennas with largest channel gains. It can be observed that the Air-ISCC performance deteriorates with the increasing number of IoT devices, since it is harder to design one common data aggregation beamformer to equalize channels of more devices. Moreover, both the optimized shared and separated designs perform better than the baseline with AS, which verifies the necessity of beamformer design. The performance of the separated design is better when the number of devices is small, as it has dedicated beamforming vector for AirComp. When the number of devices is large, the shared design performs better as it is free of interference caused by sensing signals.
  
\subsection{Target Estimation for Air-ISCC}
To evaluate the performance of target estimation, the \emph{Cramér-Rao bound} (CRB) is considered, which represents the lower bound on the variance of unbiased estimators. Specifically, the MSE of AirComp is minimized while guaranteeing a series of pre-defined levels of CRB for radar sensing with respect to each IoT device as well as the transmit power budget. To this end, the signal transmission beamformers of the IoT devices need to be jointly optimized. As illustrated in Fig.~\ref{FigSys}, the target to be estimated can be categorized as a point or a surface. Different categories of targets correspond to different formulations of design problems. 

%\begin{center}
%\noindent \fbox{\parbox{.9\linewidth} {\center \textbf{Guideline for Air-ISCC Target Estimation:} \\
%--------------------------------------------------------------------\\
%Derive CRB for different types of targets.}}
%\end{center}

\textbf{Research opportunities:}

\begin{itemize}
\item {\bf Point target estimation via Air-ISCC}: In this case, the target is regarded as a point that is far away from the IoT devices. The target response matrix can be expressed as a function of the azimuth angle between the target and the IoT device. Under such condition, the CRB for target estimation is determined by the azimuth angle and the transmission beamforming matrix. Aiming at minimizing the AirComp MSE while guaranteeing the CRB requirements, the joint optimization of transmission beamformers results in a non-convex problem. After SDR, the solving approach of rank reduction might be applied to obtain the low-rank solutions, while the optimality remains to be proved.

\item {\bf Surface target estimation via Air-ISCC}: In this case, the target is modeled as a surface with a series of distributed point-like scatterers. The target response matrix should be a function of a vector composed of the angles between the scatterers and the IoT devices. Under such condition, the echoes can be randomly reflected in each of the radar’s illuminations, and thus the IoT devices have no prior knowledge about the number and the corresponding angles of scatterers, which makes the problem more challenging to be dealt with.
\end{itemize}

\subsection{Waveform Design for Air-ISCC}
Focusing on a MIMO Air-ISCC system, several dual-functional waveform design criteria need to be considered for minimizing the AirComp MSE given the constraint of radar sensing. Both the design problems of omnidirectional and directional beampattern are investigated. For an omnidirectional beampattern, the transmit waveform matrix should be orthogonal, i.e., the corresponding covariance matrix should be an identity matrix. The corresponding problem can be regarded as an \emph{orthogonal Procrustes problem} (OPP), which has closed-form solution based on the \emph{singular value decomposition} (SVD). As for directional beampattern design, the covariance of waveform matrix is fixed as a Hermitian positive semidefinite matrix, where the optimal solution can be obtained via alternating optimization. 

%\begin{center}
%\noindent \fbox{\parbox{.9\linewidth} {\center \textbf{Guideline for Air-ISCC Waveform Design:} \\
%--------------------------------------------------------------------\\
%Balance the sensing and AirComp performance via waveform design.}}
%\end{center}

\textbf{Research opportunities:}

\begin{itemize}
\item {\bf Tradeoff between sensing and AirComp}: To balance the trade-off between the radar sensing and AirComp performance, a tolerable mismatch between the designed and the desired radar beampattern might be allowed. Therefore, a weighted optimization problem for designing dual-functional waveforms is formulated, which aims at minimizing the weighted summation of both AirComp MSE and the mismatch of radar beampattern under both total and per-antenna power constraints. The solving approaches remain to be investigated.

\item {\bf Waveform design with practical constraints}: In practical waveform design, the \emph{constant modulus constraint} (CMC) and the \emph{similarity constraint} (SC) need to be enforced, which make the optimization problems more difficult to be solved.
\end{itemize}

\begin{figure*}[ht]
\centering
\includegraphics[scale=0.45]{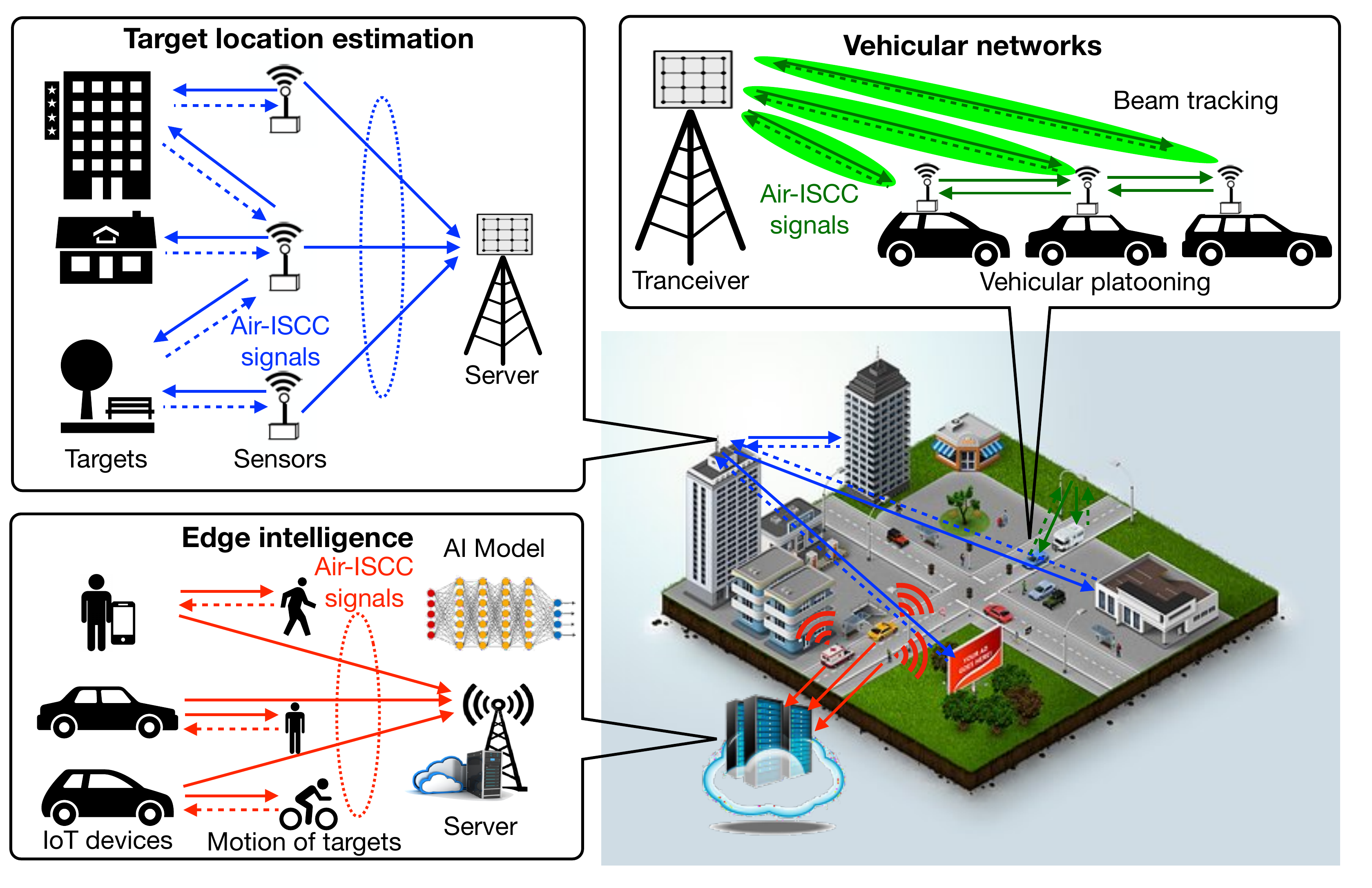}
\caption{Illustration of Air-ISCC applications.}
\label{FigApp}
\end{figure*}

\subsection{Security Design for Air-ISCC}
The IoT devices with terrible sensing capabilities or channel conditions might result in abnormal collected data, while the analog-wave addition in Air-ISCC makes it vulnerable to random or intended modifications of the sensed parameters from malicious devices, which is known as the \emph{Byzantine attacks}. Moreover, Air-ISCC faces the risk of eavesdropping, where the illegal server can also receive the computation results. To deal with the mentioned problems, secure design for Air-ISCC is needed to be investigated.

%\begin{center}
%\noindent \fbox{\parbox{.9\linewidth} {\center \textbf{Guideline for Air-ISCC Secure Design:} \\
%--------------------------------------------------------------------\\
%Improve Air-ISCC security against attacks and eavesdropping.}}
%\end{center}

\textbf{Research opportunities:}

\begin{itemize}
\item {\bf Secure Air-ISCC design against Byzantine attacks}: In separated communication and computation designs, an effective design against Byzantine attacks is to select the value minimizing the geometric median after recovering parameters from all IoT devices. As the computation is performed via transmission in Air-ISCC, new designs are required. A possible approach is to divide the IoT devices into multiple clusters with different cluster performing Air-ISCC in different time slots. The server can then calculate the geometric median after receiving the computation results from every clusters. Such an approach is expected to deal with the abnormal values caused not only by the attackers but also the devices with terrible sensing capabilities or channel conditions, while the clustering design needs to be further investigated to realize this vision.

\item {\bf Secure Air-ISCC design against eavesdropping}: To prevent eavesdropping, jammer is applied to deteriorates the quality of the received signal at the eavesdropper. However, as the eavesdroppers are usually silent with the unknown locations, jammers have to be placed at multiple locations, which will deteriorate the performance of Air-ISCC. A possible solution for this problem is to set a jammer near the legitimate server which knows the jamming sequence. Therefore, the jamming signal is strong enough at the legitimate server to be detected and canceled from the received signal. In contrast, the signal received by the eavesdropper is very similar to the white noise and thus cannot be recovered. The particular jammer location and coding methods warrant further investigation.
\end{itemize}

\section{Air-ISCC Applications}
Due to its promising performance, Air-ISCC is expected to support a wide range of IoT applications in the areas of target location estimation, vehicular networks, and edge intelligence, as illustrated in Fig. \ref{FigApp} and discussed in the following.

\subsection{Target Location Estimation}
\begin{figure*}[ht]
\centering
\includegraphics[scale=0.45]{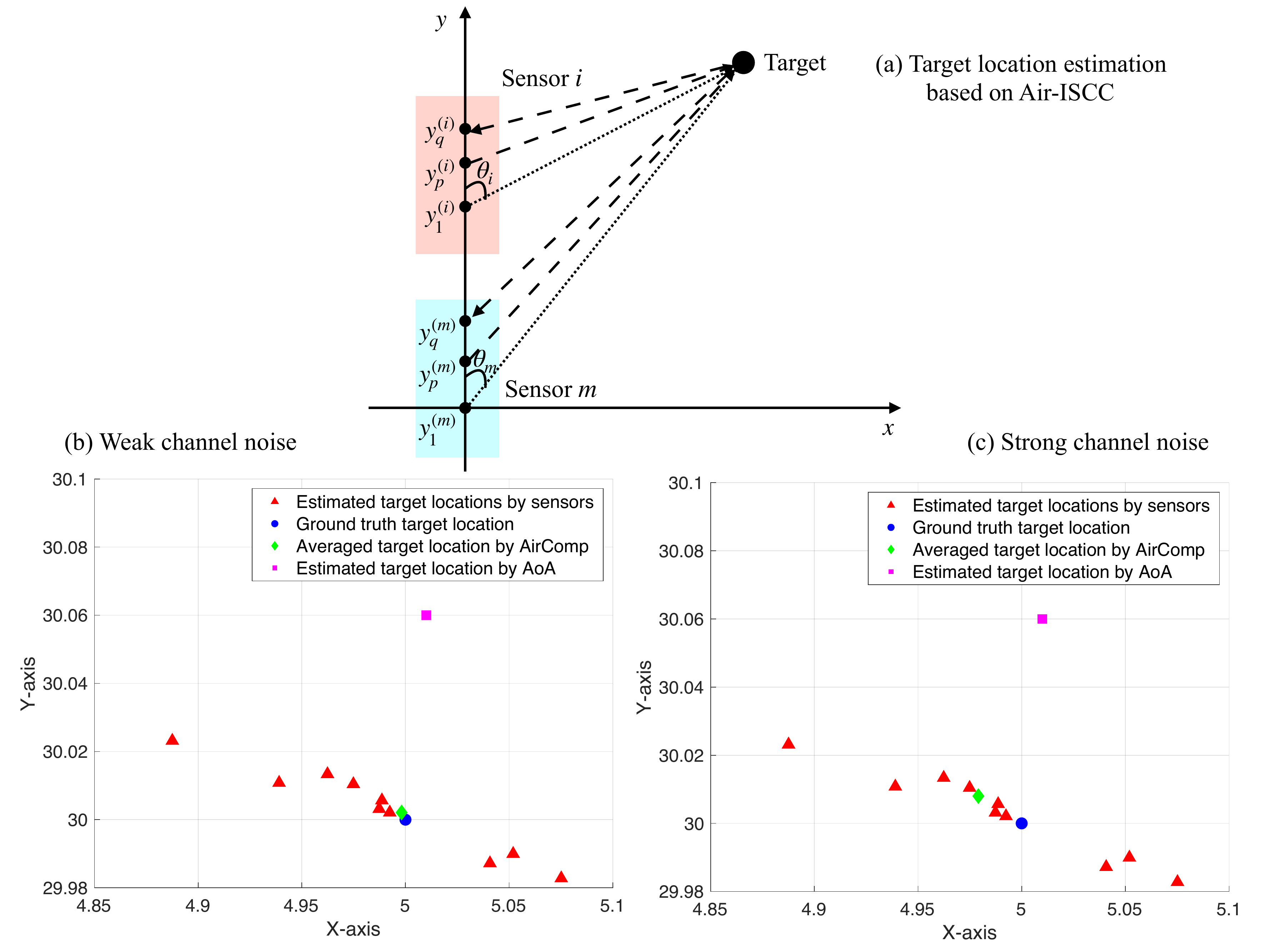}
\caption{Target location estimation based on Air-ISCC.}
\label{FigLocation}
\end{figure*}

Due to its advanced functionality, Air-ISCC can be applied for target location estimation. As illustrated in Fig. \ref{FigLocation} (a), the location of the target is estimated by multiple IoT devices based on the information of relative distance and angle estimated from the reflected radar signals as well as their own locations. Meanwhile, each IoT device delivers its previous locally estimated location of the target to the server simultaneously, and thus the server will obtain the averaged estimated target location via AirComp. Compared with the conventional communication methods, AirComp is applied to avoid the competition of radio resource among IoT devices. According to \cite{bekkerman2006target}, the azimuth angle between the target and each IoT device can be recovered from the phase delay between the two antennas at that IoT device. As it is very hard if not impossible to express the azimuth angle in closed form, the golden section search or grid search can be applied to find the numerical results. Meanwhile, the free space propagation law is adopted to estimate the distance between the target and each IoT device. Based on the estimated angle and distance, as well as its own location, each IoT device can obtain its local estimation of the target location. The target location estimated by each IoT device is then modulated (pre-processed) into data symbols. The data symbols of all IoT devices are transmitted to the server simultaneously. After AirComp and post-processing, the averaged estimated target location can be recovered at the server.

The performance of target location estimation based on Air-ISCC under the weak channel noise is illustrated in Fig. \ref{FigLocation} (b). The information to be estimated and transmitted is a vector containing the two-dimensional location of the target. The performance of the conventional radar sensing scheme based on \emph{angle of arriving} (AoA) is also illustrated, the estimated target location is obtained by minimizing the summation of mean-squared errors between the estimated angles and the actual angles over all IoT devices. One can observe that the estimated target location by each IoT device based on radar sensing is a little deviating from the ground truth, while the application of AirComp can alleviate such deviation by averaging the measured values of IoT devices over transmission. Moreover, the target location estimated by Air-ISCC is closer to the ground truth than that by the conventional AoA.

Under the strong channel noise, the performance of Air-ISCC is illustrated in Fig. \ref{FigLocation} (c). One can observe that the performance of AirComp is deteriorated due to the strong noise. In such condition, the target location estimation by a single IoT device might achieve better performance, which necessitates the device scheduling.

\subsection{Vehicular Networks}
In vehicular networks, Air-ISCC is expected to train the beam for vehicular tracking and predicting. In particular, the transmitter sends pilots to the receiver via multiple spatial beams \cite{yuan2020bayesian}. The receiver leverages different receiving beamformers to measure the quality of the received pilots and feeds the indices of the beam pair with the highest quality back to the transmitter. Through this approach, the transmitting and receiving beams are aligned with each other. After establishing the communication link, both the transmitter and receiver need to keep tracking the variation of the optimal beam pairs to guarantee the communication quality, which is known as the \emph{beam tracking}. To this end, the receiver in conventional design needs to feed back the channel information to the transmitter in every beam tracking cycles. Fortunately, Air-ISCC can effectively remove the feedback loop by mounting a radar sensor on the transmitter to estimate the channel based on echo signals and exploiting the channel reciprocity.

\begin{figure*}[ht]
\centering
\includegraphics[scale=0.4]{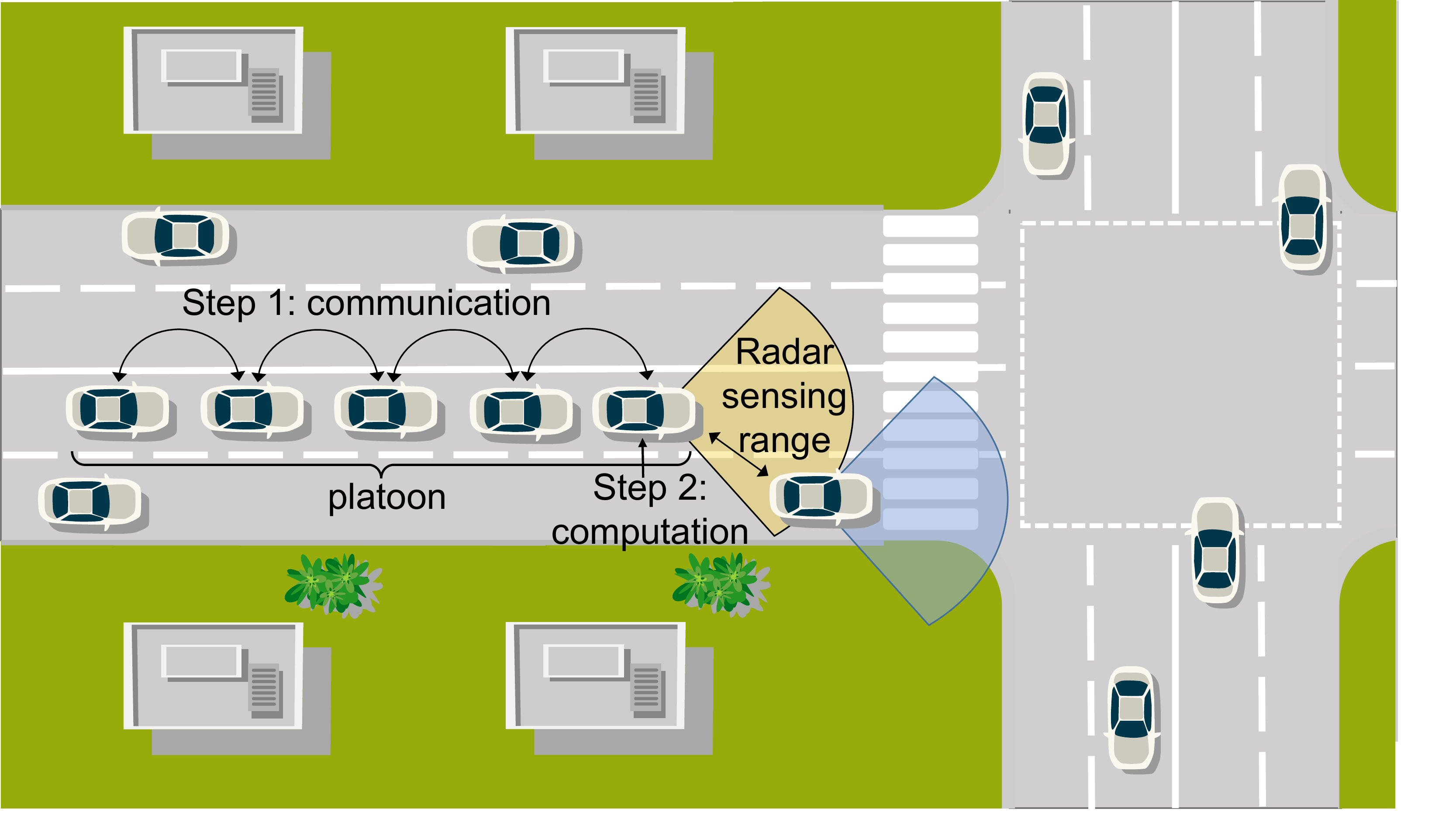}
\caption{Distributed consensus control for platooning based on Air-ISCC.}
\label{FigPlatoon}
\end{figure*}

Another application of Air-ISCC in vehicular networks is the platooning, which is achieved by the distributed consensus control. In vehicular platooning, all the vehicles in the platoon needs to reach a consensus on common driving variables including the velocity, trajectory, and acceleration \cite{jia2022online}. Therefore, each vehicle need to run an iterative consensus protocol to update its driving variables. As illustrated in Fig. \ref{FigPlatoon}, each iteration of such a protocol comprises a communication step and a computation step. The former step requires each vehicle to transmit its driving variables to other vehicles in the platoon, while in the later step each vehicle updates its driving variables according to the information received from others. The consensus is reached if all the driving variables converge to the same value. To ensure the safety of a platoon travelling at a high speed and adapting to complex traffic conditions, it is essential to guarantee its ability of responding to unforeseen events. Therefore, vehicle platooning is regarded as a mission-critical application, where the distributed consensus control process requires ultra-low latency. By enabling function computation during the concurrent transmission of driving variables from multiple vehicles, the implementation of Air-ISCC on the consensus protocol can efficiently reduce the iteration latency and thus accelerating the convergence.

\subsection{Edge Intelligence}
Aiming at gaining privacy-aware access and low-latency to distributed data for supporting the intelligent IoT, the AI tasks are spread from the centralized cloud to the wireless network edge \cite{letaief2019roadmap}. Edge intelligence has been recognized as another key technology towards the next generation wireless networks \cite{zhu2020toward}. The integration with edge intelligence is important to unlock the full potential of Air-ISCC. On one hand, Air-ISCC is expected to collect parameters such as the motion information of the sensing target at the distributed IoT devices to support model training. On the other hand, Air-ISCC enables concurrent computation and transmission of local training results from multiple IoT devices over the same frequency band for global aggregation at the server. 

In the application scenario of edge intelligence, the tradeoff between the sensing and AirComp performances in Air-ISCC is further reflected on the model training process. By improving the sensing performance at the sacrifice of the AirComp performance, the IoT devices can collect high-quality data to improve the local model training accuracy, while the quality of the wirelessly aggregated training results at the server is deteriorated due to the channel fading and noise. In contrast, to improve the aggregation performance by increasing the accuracy of AirComp, the local model training will suffer from low-quality data caused by the terrible sensing MSE. Therefore, the radar sensing and AirComp processes need to be balanced in Air-ISCC design to facilitate the development of edge intelligence.

\section{Concluding Remarks}
In the 6G wireless networks, Air-ISCC is expected to support ubiquitous and intelligent IoT services. The fundamentals of Air-ISCC are introduced in this paper, followed by the advanced Air-ISCC techniques and applications. To improve the performance of Air-ISCC, multiple potential research directions are summarized, including spatial multiplexing, target estimation, and waveform design. Air-ISCC is also expected to boost the applications of target location estimation, vehicular networks, and edge intelligence. While a partial picture was presented, we hope our discussion will spur interests and further investigations on the future evolution of wireless networks.

%\section*{Acknowledgment}
%This work is supported in part by the National Key R\&D Program of China under Grant 2019YFB1802800, National Natural Science Foundation of China under Grant 62071212 and 62101235, Guangdong Basic and Applied Basic Research Foundation under Grant 2019B1515130003, Shenzhen Science and Technology Program under Grant JCYJ20200109141414409 and JCYJ20220530113017039.

\bibliographystyle{IEEEtran}

\end{document}